\documentclass[journal=jctcce,manuscript=article]{achemso}
\author{Kuiyu Ye}
\affiliation {Key Lab of advanced optoelectronic quantum architecture and measurement (MOE), and School of Interdisciplinary Science, Beijing Institute of Technology, Beijing 100081, China}
\author{Jiale Shen}
\affiliation {Key Lab of advanced optoelectronic quantum architecture and measurement (MOE), and School of Interdisciplinary Science, Beijing Institute of Technology, Beijing 100081, China}
\author{Haitao Liu}
\affiliation{Institute of Applied Physics and Computational Mathematics, Beijing 100088, China}
\altaffiliation {National Key Laboratory of Computational Physics, Beijing 100088, China}
\author{Yuanchang Li}
\affiliation {Key Lab of advanced optoelectronic quantum architecture and measurement (MOE), and School of Interdisciplinary Science, Beijing Institute of Technology, Beijing 100081, China}
\email{yuancli@bit.edu.cn}
\author{Shengbai Zhang}
\affiliation {Department of Physics, Applied Physics, and Astronomy, Rensselaer Polytechnic Institute, Troy, New York 12180, USA}

\title[An \textsf{achemso} demo]
  {Pseudopotentials, an overlooked source and remedy of DFT errors}

\begin{document}
\begin{titlepage}






\begin{abstract}
First-principles calculations rely heavily on pseudopotentials, yet their impact on accuracy is hardly addressed. In this work, we show that most pseudopotentials to date introduce errors, which manifest themselves as errors of atomic energy levels, leading to a $de facto$ deviation from the Hohenberg-Kohn theorem. We consider the atomic-level adjusted pseudopotentials, whose interplay with exchange-correlation functional provides a pragmatic correction that balances accuracy and efficiency. We benchmark our theory with bandgap calculation for 54 semiconductors containing monovalent Cu. The results, compared to those from conventional studies, not only remove all erroneous metal predictions for 11 compounds, but also reduce the mean relative error from 80\% to 20\%. Overall accuracy even exceeds those of standard hybrid functionals and GW methods.
\end{abstract}
\end{titlepage}
\section{Introduction}

Density functional theory (DFT) has become a cornerstone in modern electronic structure calculations\cite{Jones,Huang,Lejaeghere}. By reformulating the complex interactions among electronic states in terms of total electron density, it strikes a better balance between computational efficiency and accuracy than methods solving wavefunctions directly. DFT is mostly realized by solving the Kohn-Sham (KS) equation, which uses electron density constructed from independent single-particle wavefunctions to mimic interacting many-body systems\cite{KS}. In principle, the accuracy of KS-DFT depends entirely on the accuracy of the exchange-correlation (XC) functional. However, actual accuracy also depends on the choice of numerical methods\cite{Lejaeghere}, for example, the choice of basis set can be a key factor. Various DFT codes employ planewave bases and utilize pseudopotentials to mitigate the need for very high energy cutoffs\cite{Milman,Garrity}.

Pseudopotentials predate DFT and highlight the difference between core and valence electrons\cite{Hellmann,Cohen1983}. The former are tightly bound to the nucleus and chemically inert, while the latter are more loosely bound with higher energy and chemically active. Casting the interactions of the nucleus and core electrons with valence electrons by way of pseudopotentials can greatly simplify the calculations, as it eliminates the rapid oscillations of wavefunctions near the nucleus. The pseudopotentials that are widely used today are derived from first-principles\cite{Cohen,Zunger}.

Despite a long history of research, there is still no consensus in the DFT community about the impact of pseudopotentials on accuracy. Some believe that using pseudopotentials consistent with the XC approximation is accurate\cite{Fuchs1998,Yang2023}. However, most DFT practices do not follow this. Instead, it presents a \emph{de facto} tendency for the pseudopotentials to have a negligible effect on the DFT accuracy relative to the XC functionals, as evidenced by their respective numbers. Compared to the fast-growing number of XC functionals ($>$ 500 in the LIBXC software library\cite{Lehtola}), the number of pseudopotentials is an order of magnitude smaller. As a result, in condensed matter physics, using local-density-approximation/generalized-gradient-approximation (LDA/GGA) pseudopotentials is still a common practice, leading to the ``inconsistent" scheme when advanced functionals such as meta-GGA are used\cite{Borlidojctc,Borlidonpj,Mazdziarz}. Other researchers regard pseudopotentials as effective-core-potentials for solving the KS equation and propose improving the DFT accuracy by optimizing the choice of pseudopotentials, making ``inconsistent" pseudopotential-functional calculations a natural choice. This approach has indeed been effective in calculating the bandgap, barrier height, and exciton binding energy of some notoriously DFT-failed systems\cite{TanJCP,Tanpccp,Vogel96,Vogel97,Shen2023,Shen2024,WuYJ,Bu,Rossomme,Filippetti}.

Given the above, how the pseudopotential affects the DFT accuracy, in particular its significance with respect to the XC functional? What is the hidden physics behind the better agreement with the experiment of the ``inconsistent" scheme, a necessity of the correct physics or an accident of error cancellation? If it is the former, how should we understand the relationship between pseudopotential-DFT and all-electron calculations? When implementing pseudopotential-DFT, different pseudopotential-XC combinations may all reproduce the same experimental facts. Are there any operational criteria to distinguish which is the correct result of the correct reason? These fundamental questions regarding pseudopotential-DFT remain unaddressed to this day.

In this work, we explore the answers to the above queries both theoretically and numerically. We first show that the accuracy of a pseudopotential-DFT approach is determined jointly by the pseudopotential and XC functional. As revealed in the KS equation, pseudopotentials constructed from different XC functionals carry errors of varying sizes, manifested notably in the erroneous atomic energy levels. It results in a factual deviation from the Hohenberg-Kohn theorem, which is the cornerstone of DFT. Such a serious deviation can hardly be ``corrected" by solely manipulating the XC functional. We have also clarified a misinterpretation of the consistency by revealing that its correctness relies on all-electron calculations of the exact XC rather than the approximate XC functional. We then calculate the bandgap of 54 compound semiconductors containing monovalent Cu. The conventional pseudopotential-DFT, which follow the pseudopotential-approximate XC consistency, predicted that about a quarter are metals, and more than three quarters have a bandgap at least 1 eV below experimental value. The mean relative error of the bandgap for these compounds when compared with experiment is as large as 80\%. Employing the hybrid pseudopotential, which properly corrects for the Cu atomic-level error, not only opens the bandgap for the 54 compounds mentioned above, but also reduces the mean relative error to 20\%. Such accuracy statistically outperforms even today's HSE hybrid functional and GW method, but with a marked advantage in computational efficiency.

\section{Theory of pseudopotential error}

The KS equation\cite{KS} reads
\begin{eqnarray}
	\{-\frac{1}{2}\nabla^2 + v_{nu}({\bf r}) + v_H([n];{\bf r}) + v_{xc}([n];{\bf r})\} \phi_i({\bf r}) = \varepsilon_i\phi_i({\bf r}),
\end{eqnarray}
where $[n] = \sum_i^N|\phi_i({\bf r})^2|$. $v_{nu}({\bf r})$, $v_H([n];{\bf r})$ and $v_{xc}([n];{\bf r})$ are the nuclear, Hartree and XC potentials, respectively. Inserting a potential $v_{ps}({\bf r})$ (to be determined) into Eq. (1), one obtains
\begin{eqnarray}
	\{-\frac{1}{2}\nabla^2 + v_{ps}({\bf r}) + v_H([n];{\bf r}) + v_{xc}([n];{\bf r})\} \phi_i({\bf r}) + [v_{nu}({\bf r}) - v_{ps}({\bf r})]\phi_i({\bf r}) = \varepsilon_i\phi_i({\bf r}).
\end{eqnarray}
If
\begin{equation}
	[v_{nu}({\bf r}) - v_{ps}({\bf r})]\phi_i({\bf r}) = 0 \ \ \ (|{\bf r}| > r_0),
\end{equation}
a set of wavefunctions $\psi_i({\bf r})$ can be obtained such that
\begin{eqnarray}
	\{-\frac{1}{2}\nabla^2 + v_{ps}({\bf r}) + v_H([n];{\bf r}) + v_{xc}([n];{\bf r})\} \psi_i({\bf r}) = \varepsilon_i\psi_i({\bf r}),
\end{eqnarray}
and $\psi_i({\bf r}) = \phi_i({\bf r}) \ (|{\bf r}| > r_0)$. This is exactly the equation that the pseudopotential-DFT self-consistently solves\cite{Cohen1983,Hellmann}. Therefore, any $v_{ps}({\bf r})$ satisfying Eq. (3) is a pseudopotential with $r_0$ as the cut-off radius. Physically, Eq. (3) states that at $|{\bf r}| > r_0$, the attraction of the nucleus to the valence electrons is completely cancelled out by $v_{ps}({\bf r})$, thus making them nearly-free-electrons and suitable for the planewave bases.

In principle, $v_{ps}({\bf r})$ derived from Eq. (3) is exact. However, solving Eq. (3) requires $\phi_i({\bf r})$, which comes from solving Eq. (1). Since the exact XC functional is unknown, exact $\phi_i({\bf r})$ is also inaccessible. In other words, the actual $v^{ac}_{ps}({\bf r})$, obtained under an approximate XC functional, cannot completely cancel out the nuclear attraction, leading to
\begin{equation}
	[v_{nu}({\bf r}) - v^{ac}_{ps}({\bf r})]\phi_i({\bf r}) \neq 0 \ \ \ (|{\bf r}| > r_0).
\end{equation}
This results in an error whose impact on the accuracy of the pseudopotential-DFT approach evidently also depends on the pseudopotential.

The effect of pseudopotentials on accuracy can also be visualized in a different way. Let us assume the pseudopotential in Eq. (4) is $v'_{ps}({\bf r})$,  which is different from $v_{ps}({\bf r})$. The corresponding eigen-values and eigen-functions would be $\varepsilon'_i$ and $\psi'_i({\bf r})$, respectively. Now, inserting and subtracting $v_{ps}({\bf r})$ in this modified expression, one obtains
\begin{eqnarray}
	\{-\frac{1}{2}\nabla^2 + v_{ps}({\bf r}) + v_H([n];{\bf r}) + [v_{xc}([n];{\bf r})+v'_{ps}({\bf r})-v_{ps}({\bf r})]\} \psi'_i({\bf r}) = \varepsilon'_i\psi'_i({\bf r}).
\end{eqnarray}
Note that Eq. (6) is equivalent to changing the XC functional $v_{xc}([n];{\bf r})$ to $v_{xc}([n];{\bf r})+v'_{ps}({\bf r})-v_{ps}({\bf r})$, which of course changes the electronic structure from $\varepsilon_i$ to $\varepsilon'_i$. Equation (6) shows clearly that the pseudopotential and XC functional are communicating with each other and, as such, are not completely independent from each other. It implies that a change in the pseudopotential changes the interactions between core and valence electrons. In response, it changes the XC functional form that embodies the intricate interactions among the outer-shell valence electrons.

Although somewhat interdependent, pseudopotentials and XC functionals have their own uniqueness. As such, an error due to the pseudopotential cannot be completely erased by a different choice of the XC functional. To illustrate this, let us compare Eq. (4) with Eq. (1). It is easy to see that $v_{ps}({\bf r})$ in Eq. (4) plays the same role as $v_{nu}({\bf r})$ in Eq. (1). Because $v_{nu}({\bf r})$ represents the external potential felt by electrons surrounding the nucleus, $v_{ps}({\bf r})$ can be regarded as the effective external potential for the outer-shell valence electrons. The Hohenberg-Kohn theorem states that the ground-state electron density uniquely determines the external potential. In this regard, a given $v_{ps}({\bf r})$ determines the electron density of the outer-shell valence electrons in the ground state. An incorrect $v_{ps}({\bf r})$ in Eq. (4) therefore must correspond to a defective, or even a wrong, atomic species in Eq. (1). In this sense, the pseudopotential plays a decisive role in the electronic structure, at least, on the same footing as the XC functional. Unfortunately, however, the effect of pseudopotentials has long been overlooked by the DFT community which has focused instead its attention on the development of advanced XC functionals.

A natural question, then, is how to minimize the error in solving Eq. (4). Although both the pseudopotential error [Eq. (5)] and the XC functional error arise essentially from the XC approximation, they have different physical meanings. The pseudopotential error depends on the accuracy of the XC functional in describing the screening of core electrons against the attraction between the nucleus and valence electrons, which manifests itself largely in the calculated atomic energy levels. In contrast, the XC functional error in Eq. (4) depends on the accuracy of the XC approximation in describing the interaction between outer-shell valence electrons. An XC approximation, which describes both atoms and solids well, naturally minimizes both errors. Unfortunately, however, there are almost no such functionals available. For example, the Perdew-Burke-Ernzerhof (PBE) functional\cite{Perdew1996}, which is routinely used by the condensed matter community, performs poorly when calculating atoms\cite{Zhang}. As such, when the same functional is used in constructing the pseudopotential and calculating the solid, it will inevitably produce considerable errors in one or the other. In other words, maintaining consistency between the pseudopotential and the XC approximation is not conducive to minimizing the error of Eq. (4), and the success of the aforementioned ``inconsistent" scheme is not accidental, but rather captures the physical essence.

\begin{figure}[htbp]
	\begin{center}
		\includegraphics[width=0.75\columnwidth]{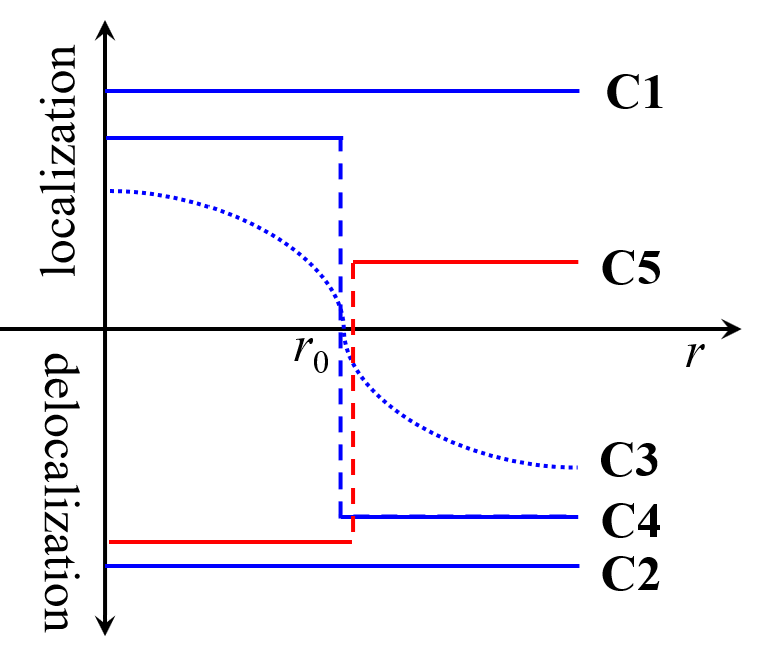}
		\caption{\label{fig:fig1}(Color online)  Schematic illustration of the difference in the consistency with respect to reproducing the all-electron calculations of the approximate XC and exact XC functionals using pseudopotential-DFT. The $r_0$ denotes the cut-off radius of the pseudopotential. Curves C1 (e.g., Hartree-Fock) and C2 (e.g., LDA/GGA) represent approximate XC functionals that are suitable for describing localized and delocalized electrons, respectively. Reproducing their all-electron calculations requires maintaining the consistency of the pseudopotential-XC approximation, i.e., the same XC functional is used on both sides of $r_0$. Curve C3 is assumed to be the exact XC, showing a transition of the electronic behavior from localized to delocalized at $r_0$ as the orbital radius increases. In the spirit of pseudopotentials, this is highly probable. Using the C1-XC to construct the pseudopotential for core electrons at $r < r_0$ while using the C2-XC to describe outer-shell valence electrons at $r > r_0$ can largely reproduce the all-electron calculations of the XC represented by the curve C4. Unambiguously, such ``inconsistent" scheme better captures the physical essence of the exact XC of the curve C3, giving higher accuracy, and is therefore the correct interpretation of the consistency. Curve C5 represents the generally accepted computational paradigm (e.g., hybrid functional), which emphasizes only the optimization of the outer-shell valence electron XC functional while maintaining the LDA/GGA pseudopotential. The dashed connection of curves C4 and C5 at $r_0$ indicates the occurrence of a jump in the localization/delocalization characteristic. For clarity, we have slightly shifted the turning point of curve C5 away from $r_0$.}
	\end{center}
\end{figure}

Then, one naturally wonders how to understand the relationship between pseudopotential-DFT and all-electron calculations. As is known, the pseudopotential was born as a mathematical tool to efficiently reproduce the results of all-electron calculations. It seems that the ``inconsistent" scheme cannot reproduce any all-electron calculation. The key lies in the difference between the exact XC and approximate XC functionals. Unambiguously, pseudopotential-DFT is intended to reproduce the results of the exact XC functional, not the approximate XC functional. It is the confusion between the two that has led to a long-standing misunderstanding of the pseudopotential-XC functional consistency in the DFT community.

We elaborate on this point with Fig. 1. The XC functionals are usually classified into two types: localized (e.g., Hartree-Fock) and delocalized (e.g., LDA/GGA), which correspond to the curves C1 and C2, respectively. In general, the larger the radius of the electron orbitals is, the more the behavior tends to be delocalized. Although the exact XC is unknown, it should be capable of correctly describing both the localized electrons of the inner shell and the delocalized electrons of the outer shell. Suppose the exact XC as curve C3, which behaves like different types of XC inside and outside $r_0$. If interpreted as the consistency of the pseudopotential with an XC approximation, what is reproduced is always the all-electron result of this localized or delocalized XC approximation. On the contrary, if using the C1-XC to construct a pseudopotential for core electrons inside $r_0$ while using the C2-XC to describe outer-shell valence electrons outside $r_0$, what is reproduced is the all-electron result of the XC represented by the curve C4. Unambiguously, the latter better captures the physical essence of the exact XC by the curve C3, giving higher accuracy. Therefore, reproducing all-electron calculations of the exact XC physically requires ``inconsistency" in the pseudopotential-XC approximation, i.e., different XC functionals are used inside and outside $r_0$. Such a consideration coincides with the underlying principle of pseudopotential, which delineates between chemically-inert core electrons and chemically-active valence electrons by a radius $r_{0}$.  Rather, enforcing pseudopotential-XC approximation consistency in fact abandons this distinction. It also means that there is a need to change the generally accepted calculation paradigm, which emphasizes only the optimization of the outer-shell valence electron XC functional while maintaining the LDA/GGA pseudopotential (e.g., the curve C5).

This situation reminds us of the subsystem functional approach proposed by Mattsson and Kohn based on differences between bulk and edge/surface states \cite{MattPRL,MattJCP}. The crux of their discussion is that there is a difference in the XC functional between a propagating state and an evanescent state. In the current case, a core state is qualitatively different from a valence state as the former is highly confined in space within $r_{0}$, while the latter outside $r_{0}$ is delocalized, suggesting one may also need different XC functionals for the two regions. This is especially true, given that the energy difference between the core and valence electrons can be an order of magnitude larger than the work function of a typical metal. This large difference might explain from a different perspective why the seemingly ``inconsistent" scheme can substantially improve accuracy\cite{TanJCP,Tanpccp,Vogel96,Vogel97,Shen2023,Shen2024,WuYJ,Bu,Rossomme,Filippetti}, since it actually uses different XC functionals for the core and valence regions, respectively.

The rationally ``inconsistent" scheme may shed fresh light on the notorious DFT bandgap problem. Due to the lack of derivative discontinuities in the LDA/GGA XC functionals, they systematically underestimate bandgaps by 30\% $\sim$ 50\%\cite{DDP,DDS}. While such a discontinuity problem can be mitigated within a generalized KS framework that includes a non-multiplicative potential operator\cite{Seidl,Perdew2017}, the computational cost becomes very high. Since the derivative discontinuity originates only from the XC functional, adjusting the pseudopotential while maintaining the LDA/GGA functionals seems impossible to fix the bandgap problem. This is, however, not the case.  According to Eq. (6), changing the pseudopotential does change the KS eigen-values, i.e., the KS bandgap. In principle, the more accurate the XC approximation, the smaller the derivative discontinuity error. For the exact XC, this error should be zero. As illustrated in Fig. 1, changing only the pseudopotentials can produce an effect amounting to the use of a more accurate XC approximation in all-electron calculations. Thus, a rationally ``inconsistent" calculation actually ``embodies" the result of a more accurate XC functional, which certainly reduces the error. If the bandgap problem of a system is entirely atomic (i.e., due to the core-electron) in origin, even LDA/GGA XC can reproduce its experimental bandgap under pseudopotential-DFT, provided the proper pseudopotentials are used.

Before closing this section, let us consider the choice of pseudopotentials. Intuitively, since pseudopotentials model pseudo-atoms, those generated by XC functionals that more accurately describe atoms will naturally be more accurate. However, in practice it is not so straightforward, and the mutual trade-off between pseudopotentials and XC functionals must be taken into account. Assuming that the traditional criterion of correct bandgap is followed, there are clearly different combinations of pseudopotentials and XC functionals that all can satisfy it. Note that Eq. (4) applies to both solids and atoms. Any pseudopotential error has to be immediately manifested in the atomic energy levels. Therefore, it is possible to identify which is the right result for the right reason and which is an error-cancellation by virtue of whether the very pseudopotential-XC functional combination gives both the correct energy levels at the atomic limit and the correct bandgap at the solid limit. Below, we will elaborate on this point using Cu monoatom and 54 monovalent-Cu compound semiconductors. These semiconductors have a wide range of applications in energy harvesting fields such as solar cells and thermoelectrics due to their suitable bandgaps. However, they are the class of semiconductors with the most serious bandgap problem, which is not well resolved even by Hubbard + $U$, hybrid functionals, and GW methods\cite{Schilfgaarde,Vidal,Bruneval,WuYJ,Shen2023,Shen2024,ZhangM}. Sometimes, different methods give conflicting conclusions.

\section{Computational Methods}
Our DFT calculations were performed using the Quantum Espresso package\cite{Giannozzi2017} with the PBE\cite{Perdew1996} functionals for valence electrons. We compare the performance of norm-conserving Vanderbilt type of PBE-pseudopotentials\cite{Hamann,SCHLIPF2015} and hybrid-pseudopotentials generated by the PBE0 functional that calculates atoms more accurately\cite{Adamo}. In order to facilitate the elucidation of the correlation between atomic energy levels and solid-state bandgap errors, we used the hybrid-pseudopotential only for Cu, which is generated by the OPIUM with the exact exchange weight fixed at 100\%\cite{Jing2018,TanJCP}. Optical bandgaps were obtained by solving the Bethe-Salpeter equation on top of DFT results using the YAMBO code\cite{Sangalli2019}.

\section{Results and discussion}
\subsection{Cu monoatom}

\begin{table}
	\caption{The 4$s$-3$d$ energy splitting $\Delta_{sd}$ of Cu monoatom by standard PBE-pseudopotential + PBE functional (PBE), PBE-pseudopotential + HSE06 functional (HSE), and hybrid-pseudopotential + PBE functional (HYPP) calculations, along with the experiment value (Exp) for comparison. All units are in eV.}
	\setlength{\tabcolsep}{3.5mm}{
		\centering
		\begin{tabular}{c c c c c }
			\hline
			\hline
			Methods          & PBE     & HSE     & HYPP          & Exp \\
			$\Delta_{sd}$   & 0.63    & 2.65    & 4.72          & 5.04       \\
			\hline
			\hline
	\end{tabular}}
\end{table}

As mentioned earlier, the pseudopotential error will be visualised in Eq. (4) solving for atomic energy levels. Here we calculate the splitting energy $\Delta_{sd}$ between the Cu 4$s$ and 3$d$ atomic states. Table I summarizes $\Delta_{sd}$ obtained with the different methods. The reason for focusing on $\Delta_{sd}$ is that it has an inherent connection with the bandgap $E_g$ of monovalent-Cu compound semiconductors. The underlying physics can be understood as follows. The Cu has a nominal electronic configuration of 3$d^{10}$4$s^1$. Upon crystallizing into a monovalent solid, the Cu loses one 4$s$ electron. Typically, the Cu 4$s$ and 3$d$ orbitals dominate the band-edge states in the conduction and valence bands. Additionally, the crystal field generated by the anionic ligand causes the Cu 3$d$ orbitals to split, which would raise the valence band maximum by $\Delta_{cf}$. Combining the two, $E_g$=$\Delta_{sd}-\Delta_{cf}$, as illustrated in Fig. 2 for Cu$_2$S.

\begin{figure}[htbp]
	\includegraphics[width=0.9\columnwidth]{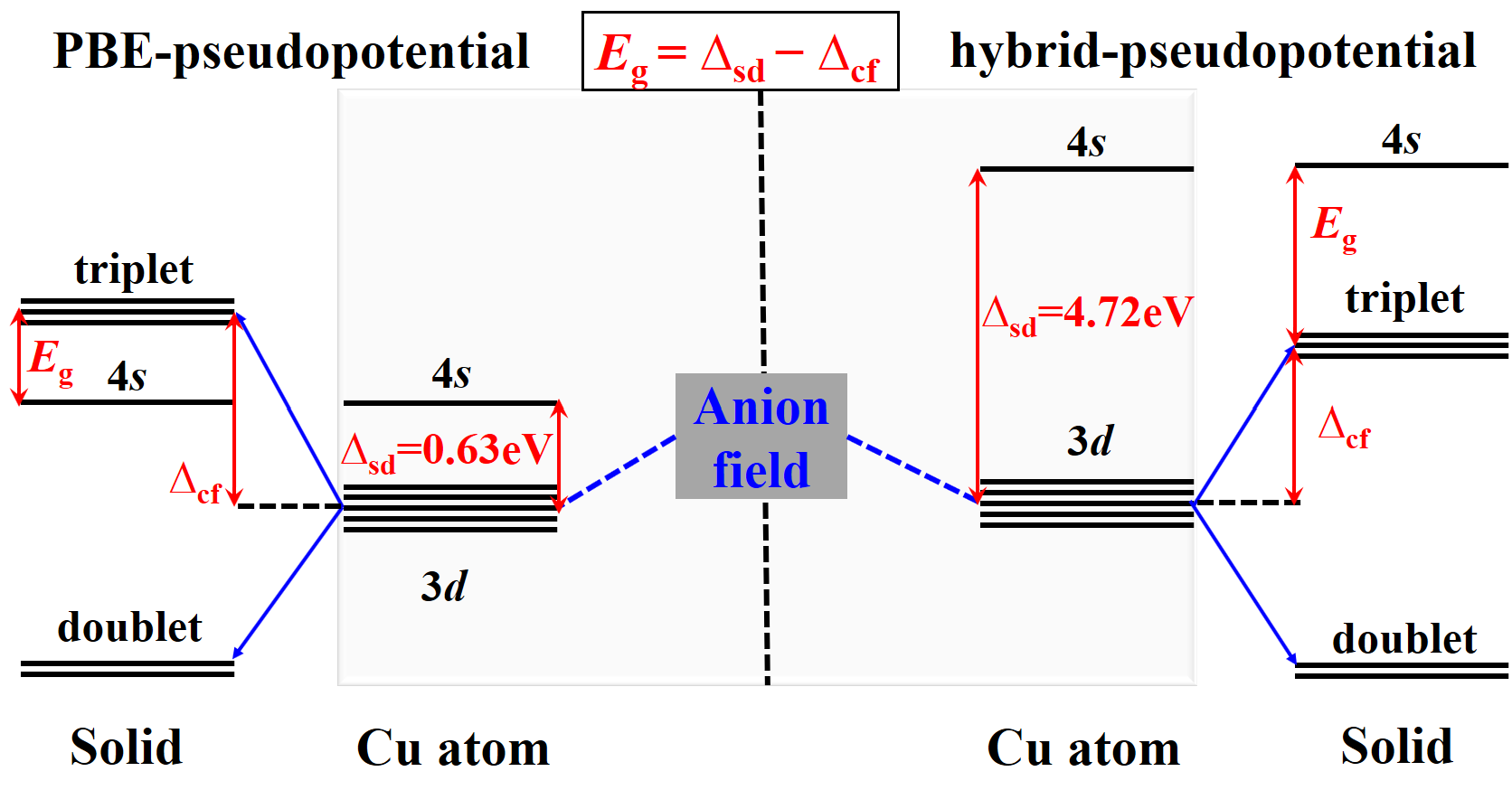}
	\caption{\label{fig:fig2} (Color online) A schematic representation of the bandap underestimation in monovalent-Cu compounds due to atomic-level errors of the pseudopotential. We compare between the PBE-pseudopotential (Left panel) and hybrid-pseudopotential (Right panel), by taking Cu$_2$S as an example. Due to the tetrahedral crystal field, the Cu-3$d$ orbitals split into a higher triplet and a lower doublet. According to our first-principles calculations, the bandgap $E_g$ can be roughly approximated as $E_g = \Delta_{sd} - \Delta_{cf}$, where $\Delta_{sd}$ is the atomic 4$s$-3$d$ splitting energy of the Cu and $\Delta_{cf}$ is the raised energy from the anionic crystal field. The significant atomic-level error of the PBE-pseudopotential causes a too small $\Delta_{sd}$. When $\Delta_{sd} < \Delta_{cf}$, a negative $E_g$ is even given on the solid side, resulting in erroneous metal predictions. On the contrary, the use of hybrid-pseudopotential largely corrects the atomic-level error, giving a close experimental $\Delta_{sd}$, thus systematically fixing the bandgap underestimation.}
\end{figure}

The left panel of Fig. 2 shows the PBE results that follow the pseudopotential-XC approximation consistency, while the right panel shows the hybrid-pseudopotential results for the rationally ``inconsistent" scheme. The PBE-pseudopotential yields a considerable error on the atomic levels with $\Delta_{sd}$=0.63 eV, which is nearly an order of magnitude smaller than the experimental value of 5.04 eV\cite{Mann}. A too small $\Delta_{sd}$ here inevitably leads to a significant underestimation of $E_g$ for the solid. When the crystal field splitting $\Delta_{cf} > \Delta_{sd}$, even a negative $E_g$ would be obtained. This is the reason why a quarter of the monovalent-Cu compounds are predicted incorrectly to be metal (see Fig. 3). Putting this in a different way, here the potential felt by outer-shell valence electrons is very different from that of a true Cu atom. In contrast, the hybrid-pseudopotential predicts $\Delta_{sd}$=4.72 eV, which is much closer to experiment 5.04 eV\cite{Mann}. As reflected by Fig. 3, this atomic difference largely fixes the solid bandgap problem. Note that with the PBE-pseudopotential, changing the XC functional to HSE06 for valence electrons cannot truly fix the problem of a too small $\Delta_{sd}$, as $\Delta_{sd}$=2.65 eV for HSE06 (see Table I) is still too small to prevent some of the erroneous metal predictions (see Fig. 4).

\subsection{Monovalent-Cu compound semiconductors}

\begin{figure*}[htbp]
	\includegraphics[width=1\columnwidth]{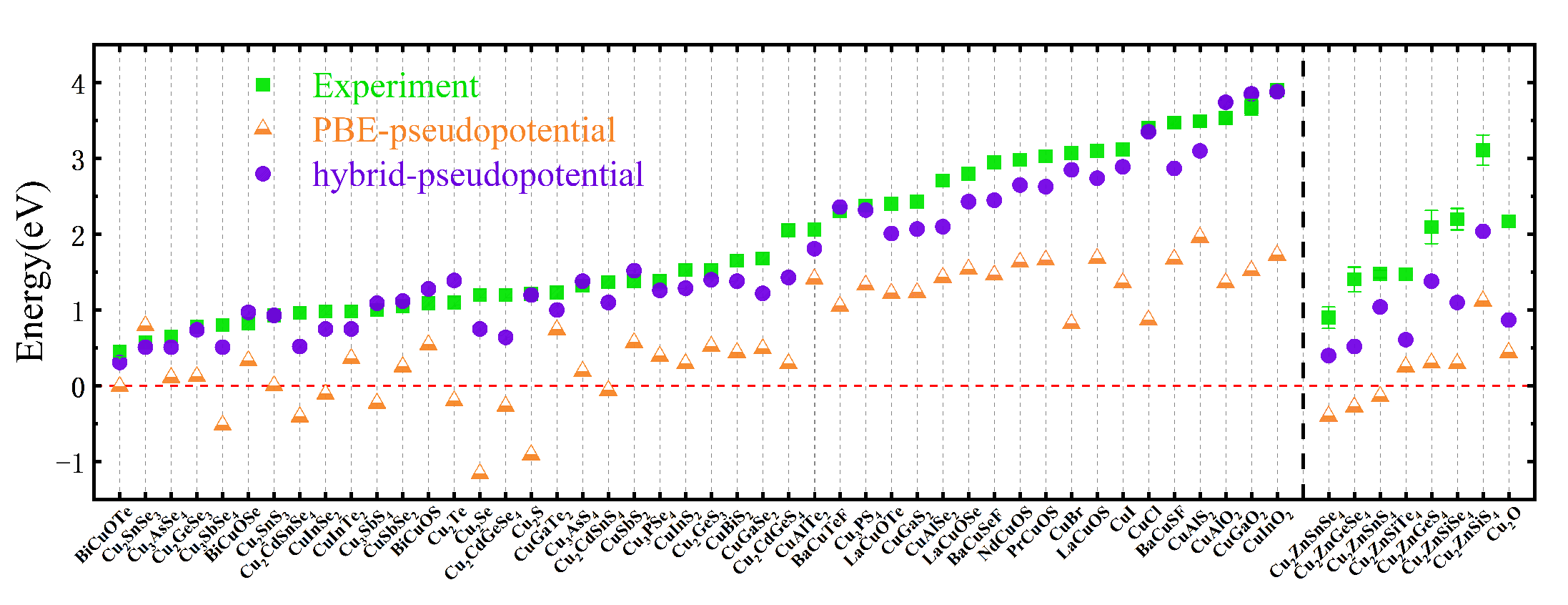}
	\caption{\label{fig:fig3} Calculated bandgaps of 54 monovalent-Cu semiconductors with the same PBE functional, but respectively using PBE-pseudopotential (orange hollow-triangles) and hybrid-pseudopotential (blue solid-balls), along with the known experimental values (green solid-rectangles). See Table S1 of the Supporting Information for details\cite{SI}. Right panel shows 8 compounds for which the bandgap problem is known to be also associated with the outer-shell valence electrons. More relevant optical gaps are employed here for three delafossite transparent conductive oxides, namely, CuAlO$_2$, CuGaO$_2$ and CuInO$_2$. In spite of a zero global gap, by convention, the Cu 4$s$ and 3$d$ energy difference at the $\Gamma$ point is defined as a negative gap.}
\end{figure*}

Indeed, when we calculate the bandgap of the monovalent-Cu solids using the atomic-level correct hybrid-pseudopotential, the accuracy with respect to the experimental values far exceeds that of the PBE-pseudopotential results, as compared in Fig. 3 (See Table S1 of the Supporting Information for details\cite{SI}). Using the PBE-pseudopotential, there are 14 species wrongly predicted to be metals, about a quarter of the total. Up to 78\% show an absolute error of more than 1 eV. Overall, the mean relative error is 80\%. The underestimation is systematic, with almost no class of Cu-compounds performing well.

Simply replacing the PBE-pseudopotential with the hybrid-pseudopotential for Cu without changing anything else largely fixes the underestimation. On the one hand, all calculated bandgaps are positive and qualitatively incorrect predictions for metals are completely eliminated. This is important as it shows that the hybrid-pseudopotential + PBE functional approach is qualitatively correct. In the DFT community, qualitative insulator-metal misprediction is usually taken as a sign that the system has strong electronic correlations and may be a potential Mott insulator. Monovalent-Cu has a $d^{10}$ electronic configuration. Typically, it forms semiconductors that do not involve partially filled $d$-electrons and should be weakly correlated\cite{Schilfgaarde}. It has therefore been a mystery as to why so many metal predictions of DFT occur. Our calculations and analysis here solve the mystery.

The hybrid-pseudopotential systematically increases the bandgaps, bringing most close to the experimental values. There are 72\% with bandgap errors less than 0.4 eV, and the mean relative error is significantly reduced to 20\%. Note that the cases considered here include 7 compounds containing both Cu and Zn bi-transition-metals. Overall, they present a more significant underestimation (see the right panel of Fig. 3), with mean absolute and relative errors of 0.79 eV and 46\%, respectively. As known, DFT also severely underestimates bandgaps of Zn compounds, and much of the underestimation stems from the inappropriate description of the outer-shell valence electrons by the PBE XC functional\cite{Borlido2019}. Therefore, solving their bandgap problem requires further corrections to the PBE XC functional. For the same reason\cite{TanJCP}, a larger error for Cu$_2$O is also expected.

\begin{figure}[htbp]
	\begin{center}
		\includegraphics[width=0.9\columnwidth]{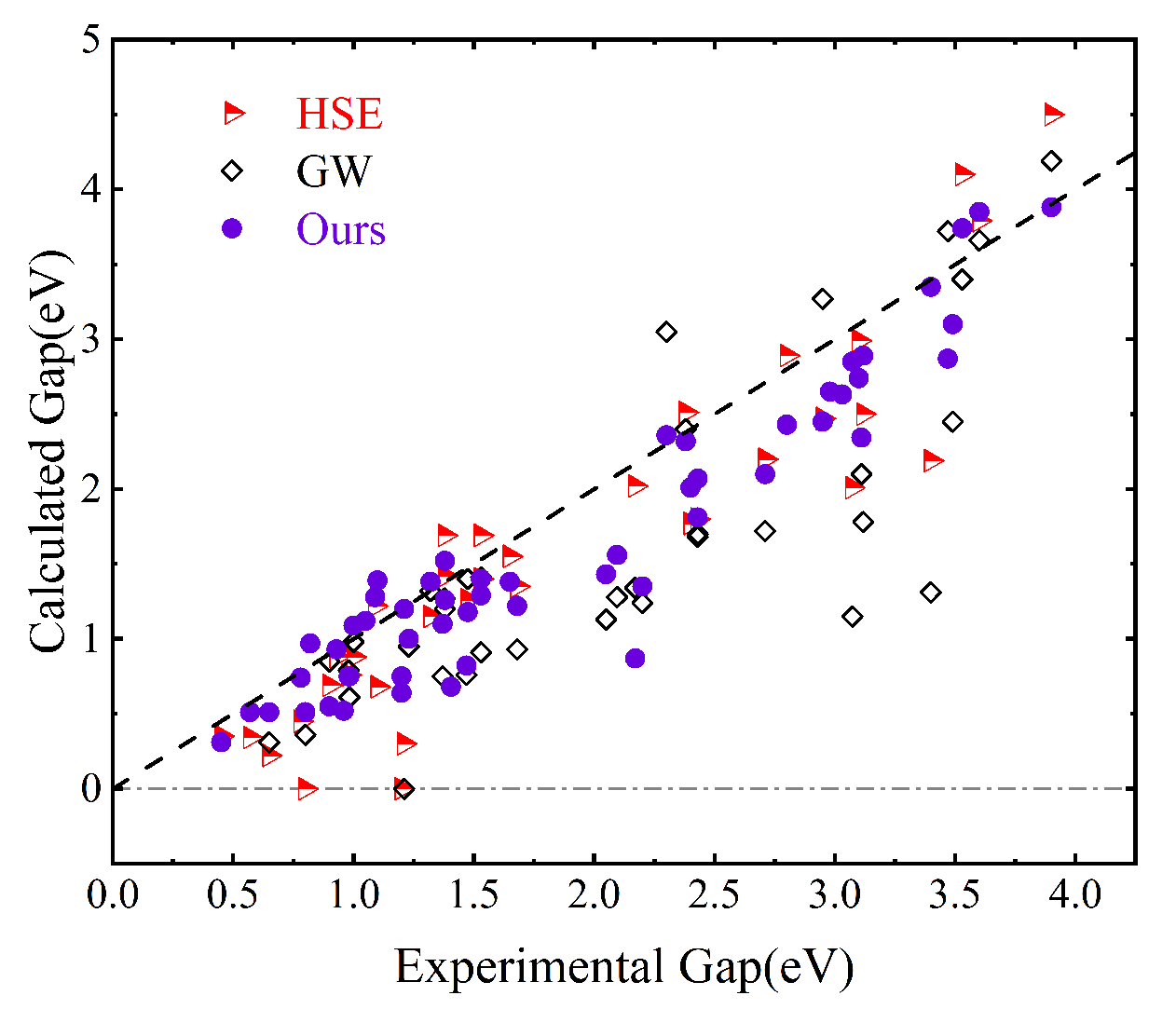}
		\caption{\label{fig:fig4}(Color online) Calculated versus experimental bandgaps for 54 monovalent-Cu semiconductors, with the dashed line indicating perfect agreement between the two. The blue solid-ball data are from our hybrid-pseudopotential calculations, and the red semi-hollow-triangle (HSE) and black hollow-diamond (GW) data are from previous studies. See Table S1 of the Supporting Information for more details\cite{SI}.}
	\end{center}
\end{figure}

In Fig. 4, we further compare the hybrid-pseudopotential results with those reported by the HSE hybrid functional and GW method (See Table S1 of the Supporting Information for details\cite{SI}). It is evident that neither HSE nor GW can avoid the erroneous metallic prediction. This proves once again that the physical essence of the bandgap problem in monovalent-Cu semiconductors lies in the pseudopotential. Also, the fact clearly supports our aforementioned assertion that the pseudopotential error can hardly be ``corrected" by solely manipulating the XC functional. Quantitatively, the mean absolute (relative) errors for the HSE and GW are 0.38 eV (25\%) and 0.58 eV (29\%), respectively, both of which are higher than 0.34 eV (20\%) observed for the hybrid-pseudopotential. In addition to the higher accuracy, the efficiency of the hybrid-pseudopotential + PBE functional calculation is also much better than that of the HSE and GW, being at the same computational level as that of KS-DFT.

It is worth emphasizing that the hybrid-pseudopotential improves the accuracy more than just the fundamental gap. Shown in Fig. 3 are the optical gaps of three delafossite transparent conducting oxides CuAlO$_2$, CuGaO$_2$ and CuInO$_2$. An exciton binding energy of 1.17 eV is deduced for CuAlO$_2$, which matches very well with recent experimental report\cite{Kang} but is more than twice as much as the value estimated on top of the PBE-pseudopotential\cite{Laskowski}. In addition, the $d$-orbital energy, effective mass and dielectric constant given by the hybrid-pseudopotential are also in agreement with experiments\cite{WuYJ,Shen2023}.

\section{Conclusion}
In conclusion, we demonstrate both theoretically and numerically that the role of the pseudopotential in electronic structure is as important as that of the XC functional, although in a different way. In general, quantum chemistry methods that perform better on atoms would produce better pseudopotentials. Our hybid-pseudopotential study on the bandgap of 54 monovalent-Cu semiconductors supports this assertion. Our findings clarify a misunderstanding of the pseudopotential-functional consistency, that is, the consistency of the pseudopotential-XC approximation is not physically necessary. Instead, the rationally ``inconsistent" scheme is the solution that captures the essence of physics, which also provides a better balance between accuracy and efficiency.

The DFT bandgap problem is routinely understood from an all-electron perspective, but most practical calculations employ pseudopotential-DFT. Our work fills the gap between the two. On the one hand, the ``inconsistent" pseudopotential-DFT in fact offers a possibility to exploit the effects of the as yet unknown XC functionals. On the other hand, the widely used LDA/GGA pseudopotentials suffer from inherent atomic energy-level errors that can lead to serious deviations from the Hohenberg-Kohn theorem. In this context, no matter how the XC functional is improved, it is impossible to produce correct results in both the atomic and solid limits. This may be one of the reasons why DFT studies sometimes yield the correct results for the wrong reasons\cite{Schiffer}. We hope that our work will evoke the DFT community to revisit the importance of pseudopotentials. Although being correct in both the atomic and solid limits is only a necessary but not a sufficient condition for obtaining physically correct results, once this is achieved it will be possible to accurately describe the entire evolutionary process of the solid formation from the atom.

\begin{acknowledgement}
This work was supported by the Ministry of Science and Technology of China (Grant No. 2023YFA1406400) and the National Natural Science Foundation of China (Grant No. 12474064). SBZ has no financial interests in any of the fundings mentioned above.
\end{acknowledgement}



\bibliography{reference.bib}

\end{document}